\documentclass[twocolumn,preprintnumbers,amsmath,amssymb]{revtex4}

\usepackage{graphicx}
\usepackage{dcolumn}
\usepackage{bm}

\begin{document}







\def\beq{\begin{equation}}
\def\eeq{\end{equation}}
\def\bea{\begin{eqnarray}}
\def\eea{\end{eqnarray}}
\def\ben{\begin{enumerate}}
\def\een{\end{enumerate}}
\def\la{\langle}
\def\ra{\rangle}
\def\a{\alpha}
\def\b{\beta}
\def\g{\gamma}\def\G{\Gamma}
\def\d{\delta}
\def\e{\epsilon}
\def\phi{\varphi}
\def\k{\kappa}
\def\l{\lambda}
\def\m{\mu}
\def\n{\nu}
\def\o{\omega}
\def\p{\pi}
\def\r{\rho}
\def\s{\sigma}
\def\t{\tau}
\def\L{{\cal L}}
\def\S{\Sigma }
\def\gsim{\; \raisebox{-.8ex}{$\stackrel{\textstyle >}{\sim}$}\;}
\def\lsim{\; \raisebox{-.8ex}{$\stackrel{\textstyle <}{\sim}$}\;}
\def\gtrsim{\gsim}
\def\lessim{\lsim}
\def\loc{{\rm local}}
\def\vm{v_{\rm max}}
\def\bh{\bar{h}}
\def\del{\partial}
\def\nab{\nabla}
\def\half{{\textstyle{\frac{1}{2}}}}
\def\fourth{{\textstyle{\frac{1}{4}}}}

\title{Energy in the Einstein-Aether Theory}

\author{Christopher Eling}
 \email{cteling@physics.umd.edu}
 \affiliation{Department of Physics, University of Maryland\\ College Park, MD 20742-4111 USA}


\begin{abstract}

We investigate the energy of a theory with a unit vector field
(the ``aether") coupled to gravity. Both the Weinberg and Einstein
type energy-momentum pseudotensors are employed. In the linearized
theory we find expressions for the energy density of the 5 wave
modes. The requirement that the modes have positive energy is then
used to constrain the theory. In the fully non-linear theory we
compute the total energy of an asymptotically flat spacetime. The
resulting energy expression is modified by the presence of the
aether due to the non-zero value of the unit vector at infinity
and its $1/r$ falloff. The question of non-linear energy
positivity is also discussed, but not resolved.

\end{abstract}

\maketitle

\section{Introduction}
\label{sec:intro}

Although Lorentz invariance has been a key feature of theoretical
physics for a century, recently there have been a number of
reasons for questioning whether it holds at all energy scales. For
example, some possible quantum gravity effects hint that it may
not be a fundamental symmetry \cite{1}. Thus, it is useful to
construct effective, low energy symmetry breaking models in the
regimes of the Standard Model and General Relativity (GR). The new
effects that appear can then be studied in a familiar context and
compared with observations. In the flat spacetime background used
in the Standard Model, Lorentz invariance can be broken by
background tensor fields \cite{2}. However, when we attempt to
couple such fields to gravity, they will also break the general
covariance of GR, which we regard as fundamental. In order to
bypass this problem, it is straightforward to consider these
fields as dynamical quantities along with the metric. In this
paper, the source of the Lorentz violation (LV) will be modeled as
a unit timelike vector $u^a$. The unit timelike restriction
preserves the well-tested $SO(3)$ group of rotations while
enforcing the breaking of boost symmetry at every point in the
curved spacetime. Therefore, the vector $u^a$ can be said to act
as an ``aether".

Similar work was initiated in the early 1970's by Will, Nordtvedt
and Hellings \cite{3,4,5} who studied a vector-tensor model
without the constraint in the context of alternative theories of
gravity. For a review of more recent work on the subject,
including aspects of observational constraints, waves, cosmology,
and black holes, see \cite{6} and the references therein.
Following these authors, we will refer to this theory as the
``Einstein-Aether" theory. One important open question is whether
the Einstein-Aether theory is energetically viable and stable. The
Will-Nordtvedt-Hellings models, for example, are unstable because
fluctuations of the unconstrained vector can be either timelike or
spacelike, allowing ghost configurations and energy of arbitrary
sign \cite{7}.

Energy in a field theory is defined as the the value of the
Hamiltonian, which acts as the generator of time translations.
Although in diffeomorphism invariant theories there is generally
no preferred notion of time (and thus energy), in asymptotically
flat spacetimes one can naturally define the ADM and Bondi
energies associated with asymptotic time translations at spatial
and null infinity respectively. The ADM and Bondi definitions for
GR have also been shown to satisfy positive energy theorems
\cite{8}.

In this paper we examine energy in the Einstein-Aether theory.
Since it has proven difficult to directly construct the
Hamiltonian for the theory, we instead consider the pseudotensor
method of studying gravitational energy. Such an approach was
first taken in a similiar context by Lee, Lightman, and Ni
\cite{9}, who derived pseudotensors for the unconstrained
vector-tensor models but did not evaluate them on solutions.
Despite the non-covariance of pseudotensors, it is known that they
give well-defined results for the spatially averaged energy
carried by waves in linearized theory and the total energy of
asymptotically flat spacetimes. In gravitational wave physics they
provide a simple and straightforward method for calculating
averaged energy densities and the energy-momentum flux radiated
away from sources. In addition, Chang, Nester, and Chen \cite{10}
have shown that the superpotential associated with every
pseudotensor corresponds to a (albeit non-covariant) quasi-local
Hamiltonian boundary term.

We first discuss the Einstein-Aether theory and then motivate and
construct its modified Weinberg pseudotensor expression. As the
calculational and consistency check we also use a Lagrangian based
method to derive the modified Einstein ``canonical" pseudotensor
and its associated superpotential. We then apply these expressions
to solutions in both the linear and non-linear regimes. In the
linearized theory we find that the Einstein and Weinberg
prescriptions give the same energy densities for the plane wave
modes derived in \cite{11}. Restricting these densities to be
positive yields constraints on the model in terms of the
coefficients of the aether part of the action. These constraints
are also compared to results obtained in the limit where the
metric and aether decouple \cite{12}. In the full non-linear
theory the Einstein-Aether superpotential is used to obtain the
total energy for an asymptotically flat spacetime. This result
agrees with \cite{13}, where the total energy in the
Einstein-Aether theory is derived via the covariant Noether charge
formalism. We conclude with a discussion of the status of positive
energy in the non-linear regime and prospects for a positive
energy theorem.

\section{Einstein-Aether action}
\label{E-A}

We can model gravity with a dynamical preferred frame using a
timelike unit vector $u^{a}$. This vector field breaks local
Lorentz invariance spontaneously in every configuration leaving
behind the 3-D rotation group as the residual symmetry. The unit
norm condition is required to avoid ghosts and also desirable
because we regard the norm as extra information beyond what is
necessary to determine the preferred frame. Taking an effective
field theory point of view we can consider the action as a
derivative expansion, subject to diffeomorphism symmetry. The
result, up to two or fewer derivatives, is
\beq S = \frac{1}{16\pi G}\int \sqrt{-g}~ L_{\rm ae} ~d^{4}x
\label{action} \eeq
where
\beq  L_{\rm ae} = -R-K^{ab}{}_{mn} \nabla_a u^m \nabla_b u^n\\ -
\lambda(g_{ab}u^a u^b - 1). \eeq
The ``kinetic" term $K^{ab}{}_{mn}$ is defined as
\beq K^{ab}{}_{mn} = c_1
g^{ab}g_{mn}+c_2\delta^{a}_{m}\delta^{b}_{n}
+c_3\delta^{a}_{n}\delta^{b}_{m} +c_4u^au^bg_{mn} \eeq
and $\lambda$ is a Lagrange multiplier enforcing the unit timelike
constraint. $R$ is the familiar Ricci scalar and the coefficients
$c_i$ in $K^{ab}{}_{mn}$ are dimensionless constants. Note that a
term proportional to $R_{ab} u^a u^b$ is not explicitly included
as it comes about as a combination of the $c_2$ and $c_3$ terms in
(\ref{action}). The metric signature is $({+}{-}{-}{-})$ and the
units are chosen so that the speed of light defined by the metric
$g_{ab}$ is unity.

The field equations from varying the action in (\ref{action})
together with a matter action with respect to $g^{ab}$ and $u^a$
are given by
\begin{eqnarray}
G_{ab} &=& T^{(u)}_{ab}+8\pi G T^{M}{}_{ab}\label{AEE}\\
\nab_a J^{a}{}_m-c_4 \dot{u}_a \nab_m u^a &=& \l u_m,
\label{ueqn}\\
g_{ab} u^a u^b &=& 1 \label{constraint}.
\end{eqnarray}
where
\beq J^a{}_{m} = K^{ab}{}_{mn} \nabla_b u^n \label{Jdef}\eeq
and
\beq \dot{u}_a = u^b \nabla_b u_a. \eeq
Here we assume that there are no aether-matter couplings in the
matter action. The aether stress tensor is given by \cite{14}
\bea T^{(u)}{}_{ab}&=&\nab_m(J_{(a}{}^m u_{b)}- J^m{}_{(a} u_{b)} 
- J_{(ab)}u^m) \nonumber\\ &&+ c_1\, \left[(\nab_m u_a)(\nab^m u_b)-(\nab_a u_m)(\nab_b
u^m) \right]\nonumber\\ &&+ c_4\, \dot{u}_a\dot{u}_b\nonumber\\
&&+\left[u_n(\nab_m J^{mn})-c_4\dot{u}^2\right]u_a u_b \nonumber\\
&&-\frac{1}{2} L_u g_{ab}, \label{aetherT}\eea
where $L_{u} = -K^{ab}{}_{mn} \nabla_a u^m \nabla_b u^n$ and
$\dot{u}^2 = \dot{u}_a \dot{u}^a$. The Lagrange multiplier $\lambda$
has been eliminated from (\ref{aetherT}) by solving for it via the
contraction of the aether field (\ref{ueqn}) with $u^a$. As we will
see below, the the form of the aether stress tensor and
Einstein-Aether Lagrangian will be important tools in derivation of
the modified Weinberg and Einstein pseudotensors.

\section{Weinberg Pseudotensor}
\label{Weinberg}

Weinberg's pseudotensor construction \cite{15} is based on the
``field theoretic" approach to GR that treats gravity as a spin-2
field on a flat background spacetime. Using Greek indices to
represent coordinate indices, we begin by writing the metric in
coordinates such that $g_{\mu\nu} = \eta_{\mu\nu}+h_{\mu\nu}$,
where $\eta_{\mu\nu}$ is the flat Minkowski metric and $h_{\mu
\nu}$ is an symmetric tensor field with the asymptotic conditions
$h_{\mu \nu} \sim O(1/r)$, $\partial_\sigma h_{\mu\nu} \sim
O(1/r^2)$, $\partial_\tau \partial_\sigma h_{\mu\nu} \sim
O(1/r^3)$. The Einstein tensor can be expanded into a series of
parts linear, quadratic, and higher order in the field variable
$h_{\mu\nu}$. Following Ch. 20 of Misner, Thorne, and Wheeler
\cite{16} the non-linear corrections to the Einstein tensor are
defined as follows
\beq 16 \pi G~ t_{\mu\nu} \equiv 2G^{(1)}_{\mu\nu}-2G_{\mu\nu},
\label{split} \eeq
where $G^{(1)}_{\mu\nu}$ and $G_{\mu\nu}$ are the linearized and
full non-linear Einstein tensors respectively. Note that this
splitting is non-unique because it depends on the coordinate
system. Since the linearized Einstein tensor is symmetric and
satisfies a linearized Bianchi identity $\partial^\mu
G^{(1)}_{\mu\nu}=0$, it can be rewritten in superpotential form
\beq 2G^{(1)}_{\mu\nu} = H_{\mu \a \nu \b}{}^{,\a \b}
\label{doublediv}\eeq
where $H_{\mu \nu \a \b}$ has the symmetries of the Riemann tensor
$H_{\mu \nu \a \b}=H_{[\mu \nu][\a \b]}=H_{\a \b \mu \nu}$ (see,
for example \cite{17}). Using (\ref{split}) and (\ref{doublediv})
the full Einstein equation becomes
\beq H_{\mu \a \nu \b}{}^{,\a \b} = 16 \pi G~(t_{\mu \nu}+T_{\mu
\nu}). \label{eesup} \eeq
Due to the symmetries of $H_{\mu \a \nu \b}$, this implies that
$\partial^\nu(t_{\mu \nu}+T_{\mu \nu})=0$. Therefore the integral
of $t_{00}+T_{00}$ over a spacelike slice is a conserved quantity.
This conserved quantity
\bea \int (t_{00}+T_{00}) ~d^3 x &=& \frac{1}{16 \pi G}\int H_{0
\a 0 \b}{}^{,\a \b} d^3x \nonumber\\
 &=& \frac{1}{16\pi G}\oint
H_{0\a0\b}{}^{,\a} n^{\b} ~d^2x, \label{energyeqn}\eea
where $n^\b$ is the unit normal to the surface at spatial
infinity, is in fact the total energy, with $t_{00}$ acting as the
energy density of the gravitational field alone.  To sharpen this
point, consider the case where the gravitational field is weak
everywhere, allowing use of the linearized theory. The leftmost
member of (\ref{energyeqn}) then gives the total matter energy,
which in this case is the total energy. The rightmost member is
insensitive to the interior volume, so replacement by arbitrary
sources and strong fields in the interior will not affect the
identification of (\ref{energyeqn}) as the total energy.

The extension to the Einstein-Aether theory is straightforward.
The metric field equations (\ref{AEE}) take the form
\beq  \widetilde{G}_{\m\n} = G_{\m\n}-T^{(u)}_{\m\n} = 8 \pi G
T_{\m\n}.\label{AEE1} \eeq
In addition to the metric, we now decompose the aether into
background and dynamical part by writing $u^\mu =
\underline{u}^\mu + v^\mu$. Unlike normal matter fields the aether
stress $T^{(u)}_{\mu\nu}$ contains linear pieces in the
perturbation $v^\mu$ due to the fact that the aether does not
vanish in the background (since it is always a unit vector). These
linear terms will modify the Weinberg pseudotensor and
superpotential. Performing the split of the modified Einstein
tensor $\widetilde{G}_{ab}$ as in (\ref{split}) we find
\beq 16 \pi G~\widetilde{t}_{\mu \nu} \equiv
2\widetilde{G}^{(1)}_{\mu \nu}-2\widetilde{G}_{\mu \nu}
\label{aeWeinberg}\eeq
where $\widetilde{G}^{(1)}_{\mu \nu}=G^{(1)}{}_{\mu
\nu}-T^{(1)(u)}{}_{\mu \nu}$. $\widetilde{G}_{\mu \nu}$ satisfies a
Bianchi identity $\nabla^\mu \widetilde{G}_{\mu \nu}=0$ if the
aether is uncoupled to the matter and if the aether field equation
(\ref{ueqn}) is satisfied. Therefore in the linearized case we can
 write
\beq 2\widetilde{G}^{(1)}{}_{\mu \nu} = \widetilde{H}_{\mu \a \nu
\b}{}^{,\a \b} \eeq
along with
\beq \widetilde{H}_{\mu \a \nu \b}{}^{,\a \b} = 16 \pi
G(\widetilde{t}_{\mu \nu}+T_{\mu \nu}). \label{eesupae} \eeq
By the same reasoning as before we could conclude that the total
energy is given by
\beq E = \frac{1}{16 \pi G}\oint \widetilde{H}_{0\a 0 \b}{}^{,\a}
n^\b d^2x.\eeq
However, unlike the GR case (\ref{eesup}), it is not clear whether
the new Weinberg superpotential $\widetilde{H}_{\mu\a \nu
\b}{}^{,\a}$ can be expressed as a local function of the fields
$h_{ab}$ and $u^{a}$ \footnote{The author thanks an anonymous
referee for pointing out this fact}. On the other hand, the
pseudotensor $\widetilde{t}_{\mu \nu}$ can be calculated directly
via the non-linear pieces of $T^{(u)}_{\mu \nu}$ and $G_{\mu \nu}$
in $\widetilde{G}_{\mu \nu}$. This will be used to compute the
linearized wave energy densities. Evaluation of the total energy as
a surface integral at spatial infinity requires a locally defined
superpotential. Since we do not have knowledge of the aether
corrections to Weinberg superpotential we shall instead consider the
Einstein superpotential, which can be derived directly from the form
of the Lagrangian. The Einstein formulation of gravitational
energy-momentum will also provide a consistency check when we
evaluate the energy density of the linearized plane wave modes.

\section{Einstein ``canonical" pseudotensor}
\label{Einstein}

The gravitational energy pseudotensor originally derived by
Einstein in 1916 shortly after his discovery of the field
equations of GR is closely related to the familiar canonical
stress tensor of matter fields in flat spacetime. In order to
derive the corresponding expression for the Einstein-Aether
theory, we use a Lagrangian approach based upon the famous work of
Noether relating symmetries to conservation laws. In flat
spacetime, invariance of a Lagrangian under global space and time
translations is associated with the conservation of
energy-momentum expressed by the conservation of the canonical
stress tensor
\beq T_{\nu}{}^\mu = \frac{\partial {\cal
L}}{\partial(\partial_\mu \psi)}
\partial_\nu\psi-\delta^\mu_{\nu} {\cal L}, \label{flatcan}\eeq
where ${\cal L} = {\cal L}(\psi,\partial \psi)$ and $\psi$
represents a general collection of fields with indices suppressed.
In the case of local symmetries, such as the diffeomorphism
invariance of the Einstein-Aether theory, the situation is more
complex. In the Appendix we review a general formalism due to
Julia and Silva \cite{18} for constructing Noether currents and
superpotentials and apply it to the Einstein-Aether theory. Using
these results we show that the pseudotensor and superpotential
have the following general form
\begin{eqnarray}
t_{\nu}{}^\mu &=& \frac{\sqrt{-g}}{16\pi G}\left(\frac{\partial
L}{\partial(\partial_\mu g_{\alpha\beta})} \partial_\nu
g_{\alpha\beta}\right. \nonumber\\&& \left. + \frac{\partial
L}{\partial(\partial_\mu u^{\alpha})}
\partial_{\nu} u^{\alpha} - \delta^\mu_\nu
L\right) \label{Epseudo}\\
U_{\nu}{}^{\mu\gamma} &=& \frac{\sqrt{-g}}{16\pi
G}\left(\frac{\partial L}{\partial(\partial_\mu
g_{\a\b})}(\delta^{\gamma}_{\alpha}g_{\nu\beta}+\delta^{\gamma}_{\beta}g_{\nu\alpha})\right.
\nonumber\\&& \left.-\frac{\partial L}{\partial(\partial_\mu
u^{\alpha})} \delta^{\nu}_{\alpha} u^{\gamma}\right)
\label{Esuper}
\end{eqnarray}
where $L$ is the Lagrangian
\bea L &=& -
g^{\alpha\beta}(\Gamma^{\eta}_{\alpha\delta}\Gamma^{\delta}_{\eta\beta}
-\Gamma^{\eta}_{\eta\delta}\Gamma^{\delta}_{\alpha\beta})\nonumber\\&&-K^{\a\b}{}_{\m\n}
\nabla_\a u^\mu \nabla_\b u^\nu - \lambda(g_{\mu\nu}u^\mu
u^\nu-1). \eea
Note that we have eliminated a surface term in the
Einstein-Hilbert Lagrangian, replacing the Ricci scalar $R$ with
the Einstein-Schrodinger ``$\Gamma^2$'' action, which depends only
on the metric and its first derivatives. When evaluated on-shell
the pseudotensor and superpotential obey the following relations
\begin{eqnarray}
\partial_\mu t_\nu{}^\mu &=& 0 \label{divt}\\
t_\nu{}^\mu &=& -\partial_\gamma U_\nu{}^{\gamma\mu}\label{tsup}.
\end{eqnarray}
To account for the presence of any non-aether matter sources one
only has to make the replacement $t_\nu{}^\mu \rightarrow
t_\nu{}^\mu + T_\nu{}^\mu$ in (\ref{divt}) and (\ref{tsup}). Like
the Weinberg construction, the pseudotensor $t_\nu{}^\mu$ is a
conserved quantity and is related to the divergence of a
superpotential.

The contributions from the pure GR $\Gamma^2$ Lagrangian are the
{\it Einstein} pseudotensor
\begin{eqnarray} ^{\rm einstein}t^{\ \m}_\r &=& \frac{\sqrt{-g}}{16\pi G} \left(\d^{\
\m}_\r (\G^\a_{\b\g}\G^\b_{\a\d}-\G^\a_{\a\b}\G^\b_{\g\d})g^{\g\d}
\right. \nonumber\\&&
\left.+\G^\b_{\r\a}\G^\g_{\g\b}g^{\m\a}-\G^\b_{\r\b}\G^\g_{\g\a}g^{\m\a}
+\G^\a_{\r\a}\G^\m_{\b\g}g^{\b\g}\right. \nonumber\\&&
\left.+\G^\m_{\r\a}\G^\b_{\b\g}g^{\a\g} -2
\G^\m_{\a\b}\G^\a_{\r\g}g^{\b\g}\right) \label{Einsteinpsu}
\end{eqnarray}
and the {\it von Freud} superpotential (see, e.g. \cite{19})
\beq ^{\rm fr}U_{\beta}{}^{\lambda\alpha} = \frac{1}{16\pi G}
\frac{1}{\sqrt{-g}}
g_{\beta\tau}\partial_\gamma\{(-g)(g^{\lambda\tau}g^{\a\gamma}-
g^{\alpha\tau}g^{\lambda\gamma})\}. \label{Einsteinsup}\eeq
To compute the additional aether modifications, we use the
relation
\beq \frac{\partial L}{\partial(\partial_\mu g_{\a \b})} = \half
(g^{\a \nu}\delta^{\b}_{\g} \delta^{\mu}_{\d}+g^{\a \nu}
\delta^{\b}_{\d} \delta^{\mu}_{\g}-g^{\mu \nu} \delta^{\a}_{\g}
\delta^{\b}_{\d}) \frac{\partial L}{\partial(\Gamma^{\nu}_{\g
\d})} \eeq
and $L_u=K^{\a\b}{}_{\mu\nu} \nabla_\a u^\mu \nabla_\b u^\nu$ in
(\ref{Epseudo}) and (\ref{Esuper}) since when evaluated on
solutions any terms related to the unit constraint will vanish. We
find the pseudotensor
\bea ^{\rm \ae}t_{\nu}{}^{\lambda} &=& \frac{1}{16\pi G}\left(2
{\sqrt{-g}} J^{\lambda}{}_{\rho}\nabla_{\nu} u^{\rho}
-\sqrt{-g}\{(J^{\lambda}{}_{\beta}+J_{\beta}{}^{\lambda})u^{\alpha}
\right. \nonumber\\
&&
\left.-(J^{\alpha}{}_{\beta}+J_{\beta}{}^{\alpha})u^{\lambda}+(J^{\lambda\alpha}
-J^{\alpha\lambda})u_{\beta}\}
\Gamma^{\beta}{}_{\alpha\nu}\right. \nonumber\\
&& \left.+ \delta_{\nu}^{\lambda} {\sqrt{-g}} L_{u}\right).
\label{aepsu} \eea
and the superpotential
\bea ^{\rm \ae}U_\beta{}^{\lambda\alpha} &=& \frac{1}{16 \pi
G}\sqrt{-g}
\left((J^{\lambda}{}_{\beta}+J_{\beta}{}^{\lambda})u^{\alpha}
\right. \nonumber\\
&& \left.
-(J^{\alpha}{}_{\beta}+J_{\beta}{}^{\alpha})u^{\lambda}+(J^{\lambda\alpha}
-J^{\alpha\lambda})u_{\beta}\right) \label{aesup}\eea
where $J^a{}_\b$ is defined in (\ref{Jdef}). The above
decompositions of (\ref{Epseudo}) and (\ref{Esuper}) into GR and
aether pieces do not satisfy (\ref{divt}) and (\ref{tsup})
independently. A key requirement when evaluating these
pseudotensorial expressions is that the metric must be written in a
coordinate system where the connection coefficients vanish like
$O(1/r)$ or faster in the asymptotic limit. If the coordinate system
is not chosen properly then these expressions will yield incorrect
energies and momenta \footnote{For example, if one uses the
Schwarzschild metric in spherical polar coordinates $ds^2 =
(1-\frac{2M}{r})dt^2-(1-\frac{2M}{r})^{-1}dr^2-r^2 d\Omega^2$, the
von-Freud superpotential will yield an incorrect total energy. After
re-expressing the metric in Cartesian coordinates $(t,x,y,z)$, the
pseudotensor expression gives $E=M$.}. This condition was not well
understood in the early literature on gravitational energy-momentum,
but can now be explained using an analysis of the boundary terms and
conditions in an action. See the Appendix for further details.

\section{Energy in Linearized Theory}
\label{Linear}

Equipped with the modified Einstein and Weinberg pseudotensors we
can now calculate the energy density of the linearized plane wave
solutions to the Einstein-Aether theory. The plane wave solutions
in the absence of matter are found by linearizing the field
equations above, (\ref{AEE})-(\ref{constraint}), with $g_{\mu\nu}
= \eta_{\mu\nu}+h_{\mu\nu}$ and $u^\mu = \underline{u}^\mu+v^\mu$.
This gives
\begin{eqnarray}
\partial_\a J^{(1)\a}{}_\b &=& \lambda \underline{u}_{\b}\label{linaefield}\\
G^{(1)}_{\a\b} &=& T^{(1)}_{\a\b} \label{linaemetric}\\
v^0 &=& -\frac{1}{2}h_{00}\label{linaeconstr}.
\end{eqnarray}
Cartesian coordinates are used in the flat background,
$\eta_{\m\n} = (1,-1,-1,-1)$ and $\underline{u}^\m = (1,0,0,0)$.
Since the background value of the Lagrange multiplier vanishes,
$\lambda$ in (\ref{linaefield}) represents a perturbation. The
superscript (1) represents quantities written to first order in
the perturbation. Jacobson and Mattingly \cite{11} then proceed to
analyze these equations using the gauge choice
\begin{eqnarray}
h_{0i} &=& 0 \label{gauge1}\\
v_{i,i} &=& 0 \label{gauge2}
\end{eqnarray}
which they prove to be accessible. Inserting plane wave solutions
\begin{eqnarray}
h_{\mu\nu} &=& \epsilon_{\mu\nu}e^{ik_c x^c}\\
v^{\mu} &=& \epsilon^{\mu}e^{ik_c x^c}
\end{eqnarray}
into the equations of motion, imposing the 4 gauge conditions
(\ref{gauge1})-(\ref{gauge2}), and choosing coordinates such that
the wave-vector is $(k_0,0,0,k_3)$ (travelling in the $z$
direction), it is found \cite{11} that the mode polarizations and
speeds are completely determined. The result is a total of 5 wave
modes falling into spin 2, spin 1, and spin 0 types as shown in
Table \ref{speeds} \footnote{Unlike (for example) the Lorentz gauge
in GR, the residual gauge of (\ref{gauge1})-(\ref{gauge2}) is not
compatible with the equations of motion
(\ref{linaefield})-(\ref{linaeconstr}) so it is not clear how to fix
the remaining gauge in a simple way. However, it is possible to
argue for the existence of 5 wave modes by counting gauge
inequivalent degrees of freedom. Consider a theory with $N$ field
variables and $M$ gauge symmetries. One can always use the $M$
constraint equations to solve for the $M$ variables whose time
derivative does not appear in the equations of motion. This reduces
the number of degrees of freedom to $N-M$. Then the remaining $M$
gauge functions can be used to further reduce to $N-2M$. In the case
of the aether theory the metric has 10 degrees of freedom and the
constrained vector has 3, making 13 field variables. There are 4
diffeomorphism symmetries, and 13 - 2x4 = 5.}.
\begin{table*}
\caption{\label{speeds}Wave Mode Speeds and Polarizations}
\begin{ruledtabular}
\begin{tabular}{lll}
Mode&Squared Speed $s^2$ &Polarizations\\
\hline
spin-2&$1/(1-c_{13})$& $h_{12}$,$h_{11}=h_{22}$\\
spin-1&$(c_1-\frac{1}{2}c_1^2+\frac{1}{2}c_3^2)/c_{14}(1-c_{13})$&
$h_{I3}=[c_{13}/(1-c_{13})s]v_{I}$ \\
spin-0&$c_{123}(2-c_{14})/c_{14}(1-c_{13})(2+c_{13}+3c_2)$&$h_{00}=-2v_0,$\\
& &$h_{11}=h_{22}=-c_{14}v_0,$\\
& &$h_{33}=[2c_{14}(c_2+1)/c_{123}]v_0$\\

\end{tabular}
\end{ruledtabular}
\end{table*}
The notation $I$ in subscript refers to the transverse components
of the metric and aether while $c_{14}=c_1+c_4$, etc. The 2 spin-2
TT metric modes look exactly like the usual GR case, except for
the modification of the speed. The 2 spin-1 transverse aether
modes and 1 spin-0 trace mode are new modes coming from the
constrained aether, which is characterized by 3 degrees of
freedom.

In order to determine the energy, note that in the absence of
matter the Weinberg prescription (\ref{energyeqn}) reduces to
\beq E = \frac{1}{16\pi G} \oint \widetilde{H}_{0\a0\b}{}^{,\a}
n^\b ~ d^2x = \int \widetilde{t}_{00} ~ d^3 x, \eeq
which clearly produces infinite total energy for plane wave modes.
One could reformulate the problem in terms of wavepackets with the
appropriate asymptotic fall-off conditions, but a far more direct
approach is to simply evaluate the plane wave energy density
$\widetilde{t}_{00}$. This quantity is meaningless at a point for
plane waves, but the average over a cycle is well-defined.
Consider a large, but finite region with nearly plane waves. There
are ``surface effects", but the the contribution to $\int
\widetilde{t}_{00} d^3 x$ is dominated by the volume. Thus,
$\widetilde{t}_{00}$ gives an effective energy density.

The 3 general classes of modes were analyzed separately using the
Riemann tensor package \cite{20} in Maple. The package allows the
user to enter the components of the metric and aether vector,
calculate curvature tensors, and to define new tensors involving
both ordinary and covariant derivatives. In this case we entered
the linearized metric and aether, where $h_{\m\n}$ and $v^\m$ take
the plane wave forms. A polarization was written as
\beq A \exp{ik_3(z-st)} + \overline{A} \exp(-ik_3(z-st)) \eeq
where $s$ are speeds shown in Table \ref{speeds} and $A$ is a
complex-valued function. Using this metric we calculated the
explicit form of the Weinberg pseudotensor $\widetilde{t}_{00}$
(\ref{aeWeinberg}) up to quadratic order. Higher order terms will be
small in the linearized theory and oscillatory terms proportional to
$\overline{A}^2$ and $A^2$ can be neglected in the usual time
averaging process. These energy densities were then compared with
the modified Einstein pseudotensor~ $^{\rm
einstein}t_{0}{}^{0}+^{\rm \ae}t_{0}{}^{0}$ from (\ref{Einsteinpsu})
and (\ref{aepsu}) again up to quadratic order in the perturbations.
Note that while (\ref{divt}) holds at quadratic order when the
linearized equations of motion are imposed, (\ref{tsup}) does not.
Therefore, one must use the modified Einstein pseudotensor directly
to compute the energy densities. The results of the Weinberg and
Einstein prescriptions agreed and are displayed below:
\begin{eqnarray}
\mathcal{E}_{\rm spin-2} &=& \frac{1}{8\pi G}~ k_3^2 ~|A|^2\label{TT}\\
\mathcal{E}_{\rm spin-1} &=& \frac{1}{8\pi G}~ k_3^2~
|A|^2\frac{c_3^2-c_1^2+ 2c_1}{1-c_1-c_3}\label{TA}\\
\mathcal{E}_{\rm spin-0} &=& \frac{1}{8\pi G} ~k_3^2 ~|A|^2
c_{14}(2-c_{14})\label{trace}
\end{eqnarray}
These results have been independently verified in \cite{21} using
the Noether charge method and a decomposition of $u^a$ into
irreducible pieces. The lack of $c_i$ dependence in (\ref{TT}) and
the simplicity of (\ref{TA})-(\ref{trace}) is striking considering
the complicated form of the pseudotensor expressions. The energy
of the spin-2 mode is positive definite, like pure GR, while for
the other 2 modes the sign of the energy density depends upon a
combination of $c_1$, $c_3$, and $c_4$. Note that when the $c_i$'s
are zero (\ref{TA}) and (\ref{trace}) are zero as expected. This
set of results for the coefficients also holds for exponentially
growing modes (i.e. when $s^2<0$)
\beq A \cos(k z+\phi)\exp(k s t) \eeq
when we average over the spatial oscillations. Restricting $s^2
> 0$ in Table \ref{speeds} to eliminate the unstable modes and enforcing
positivity in the energy densities in (\ref{TT})-(\ref{trace})
restricts the $c_i$ values in the Einstein-Aether theory.

In \cite{12}, Lim worked in the limit where the aether and metric
perturbations decouple, with the aether propagating in flat
spacetime. Mathematically this amounts to tuning $c_i, G
\rightarrow 0$ while holding the ratio $c_i/G$ fixed in the action
(\ref{action}). If we then expand the metric as $g = \eta +
\sqrt{G}~ h$ and take the limit, the action reduces to that of
linearized gravity plus aether terms coupled only to $\eta_{ab}$.
In this limit the linearized constraint reduces to $v^0 = 0$ and
we can decompose $v^i$ into spin-0 and spin-1 parts via $v^i =
\partial^i S + N^i$ where $N^i_{,i} = 0$. By examining the
Hamiltonian of these modes they found $c_1 > 0$ for positivity in
both cases, neglecting $c_4$. We can make contact with this result
simply by examining in the small $c_i$ limit of the wave
solutions. The trace and transverse aether energy waves then
correspond to the flat spacetime spin-0 and spin-1 modes. To
lowest order in $c_i/G$ we find that
\begin{eqnarray}
c_{14} > 0\\
c_{1} > 0 \label{smallc}
\end{eqnarray}
for positive energy densities of the spin-0 and spin-1 modes
respectively. Restoring $c_4$ in the flat spacetime analysis
yields complete agreement. Note that for small $c_i$ the $s^2>0$
criteria for stable, non-exponentially growing modes reduce to
$c_1/c_{14}\geq 0$ for the spin 1 aether-metric mode and
$c_{123}/c_{14} \geq 0$ for the spin 0 trace mode. Thus, modes
with positive energy are stable if $c_{123}>0$.

\section{Non-Linear Energy}
\label{Non-linear}

In this section we will attempt to extend the criteria for
positive energy from linearized theory into the non-linear regime.
As a first step, let us consider the total energy of an
asymptotically flat spacetime in the full non-linear theory.
Integrating (\ref{tsup}) over a spacelike slice in the presence of
non-aether matter gives the total energy
\beq \int T_{\rm eff}{}_0{}^{0} = \int \partial_\lambda ~^{\rm
tot}U_0{}^{\lambda0} =  \oint_{\infty} ~^{\rm tot}U_0{}^{\lambda0}
n_\lambda dS \label{ADM} \eeq
where $^{\rm tot}U =~  ^{\rm \ae}U+~^{\rm gr}U$ are the aether and
von-Freud superpotentials, (\ref{aesup}) and (\ref{Einsteinsup}),
and $T_{\rm eff} = t+T$ is total matter and gravitational
energy-momentum. The problem now is to calculate the
superpotentials for the asymptotically flat solutions to the
Einstein-Aether theory. We will use Cartesian coordinates
throughout since these have the required asymptotic behavior
discussed at the end of Section \ref{Einstein}. Therefore, the
surface element is $dS = r^2 d\Omega^2$ and the unit normal is
$(\sqrt{2},x/r,y/r,z/r)$ where $r=\sqrt{x^2+y^2+z^2}$. For
asymptotically flat boundary conditions we will assume that as
$r\rightarrow \infty$
\bea g_{\m\n} = \eta_{\m\n}+O(1/r)+ \cdots\\
u^\m = \underline{u}^\m + O(1/r) + \cdots \eea
where $\underline{u}^\m=(1,0,0,0)$ with respect to the Minkowksi
metric $\eta_{\m\n} = (1,-1,-1,-1)$. Equation (\ref{ADM}) will
only be affected by terms in the metric and aether up to $O(1/r)$.
Using the analysis of the Newtonian limit \cite{22} and applying
the unit constraint, we find that far from the source in any
asymptotically flat solution
\begin{eqnarray}
g_{00} &=& 1-\frac{r_0}{r}+\cdots \label{g00}\\
g_{ij} &=& -1-\frac{r_0}{r}+\cdots \label{gij}\\
g_{0i} &=& O(1/r^2)+\cdots \label{g0i}\\
u^{t}  &=& 1+\frac{r_0}{2r}+\cdots \label{timecomp}\\
u^{i}  &=& O(1/r^2)+\cdots \label{spacecomp}.
\end{eqnarray}
The constant value at infinity and $1/r$ fall-off term in the
aether are due to the unit timelike constraint. Thus, unlike
ordinary fields, the aether will contribute to the energy
expression directly. Inserting (\ref{g00})-(\ref{spacecomp}) into
the von-Freud superpotential (\ref{Einsteinsup}) and aether
superpotential (\ref{aesup}) yields the usual `ADM mass'' of GR
\beq E_{\rm GR} = \frac{1}{16\pi G} \oint_{\infty}
(g_{jk,k}-g_{kk,j}) n_j d^2 S = \frac{r_0}{2G} \eeq
and the aether modification
\beq E_{\rm \ae} = \frac{c_{14}}{8\pi G} \oint_{\infty}
\partial^i u^{t}~ n_i d^2 S  = -\frac{c_{14}}{2} \frac{r_0}{2G}. \label{aeADM} \eeq
Combining, we find
\beq E_{\rm tot} = \frac{r_0}{2G}(1-\frac{c_{14}}{2})
\label{aetot}.\eeq
This shows that the aether contribution effectively renormalizes
the $r_0/2G$ value we usually find for the total energy of an
asymptotically flat spacetime in GR. This renormalization can also
be understood as a rescaling of Newton's constant of the form
$G_N=G/(1-c_{14}/2)$, which agrees with the result of \cite{22}.

Equation (\ref{aetot}) implies that if $c_{14}<2$ then the total
energy of the Einstein-Aether theory is positive if the ADM mass
$r_0/2G$ is positive. However, the positive energy theorem for GR
\cite{8} requires a stress-tensor that satisfies the dominant
energy condition. The aether stress-tensor (\ref{aetherT}) does
not appear to generally satisfy this condition, so proof of total
positive energy remains elusive. For some speculative thoughts on
modifying the positive energy theorem, see \cite{6}.

Despite these difficulties, there are special cases of the
non-linear theory that are simple enough for calculations of the
energy, yet still give important results. One sector of interest
is the non-linear decoupled limit. As discussed above in Section
\ref{Linear} this limit allows one to essentially replace $g_{ab}$
with the flat Minkowski metric $\eta_{ab}$ in the aether parts of
(\ref{action}). One significant example is $c_2=c_3=c_4=0, c_1 \ne
0$ theory. In this case the Lagrangian density for the aether is
\beq L = c_1 \eta^{ab} \eta_{mn}~ \partial_a u^m \partial _b u^n +
\lambda(u^2-1) \eeq
This corresponds to a nonlinear sigma model on the unit
hyperboloid, which has a stress tensor satisfying the dominant
energy condition. A simple way to see this is to note that the
derivatives of the individual scalar components $u^{\mu}$ and are
contracted with $\eta_{\mu \nu}$, which is positive definite on
the unit hyperboloid. Returning to the linearized plane wave
energy densities of Section \ref{Linear} we see that in this
special case of (\ref{TT})-(\ref{trace}), if $0 < c_{1} < 1$
energy is positive in both the linearized and decoupled non-linear
regimes of the theory.

Another important application of the decoupling limit relevant for
our analysis of energy is the work of Clayton \cite{23}. Clayton
examined the Maxwell-like simplified theory where $c_1 = -c_3,
~c_2=c_4=0$ in the decoupled version of non-linear Lagrangian
(\ref{action}), yielding
\beq L = \int d^3x \{\half(\partial_t u_i-\partial_i
u_0)^2-\fourth F_{ij}^2+\half
\lambda(u_0^2-\overrightarrow{u}^2-1)\}. \eeq
where $F_{\m\n} = \partial_\m u_\n - \partial_\n u_\m$. The
standard calculation of the Hamiltonian and the constraint
equations then produces the following on-shell value for the
Hamiltonian
\beq H = \int d^3x \{\half \overrightarrow{P}^2 + P^i\partial_i
u_0 +\fourth (F_{ij})^2\}. \eeq
Unlike the electromagnetic case, the second term cannot be turned
into a total divergence since now $\nabla \cdot \overrightarrow{P}
= -\lambda u_0$ on-shell. This implies that for some solutions the
value of the Hamiltonian is negative. For example, as initial data
choose $u_i$ to be the gradient of a scalar field and
$P_i=-\partial_i u_0$. Evaluating the Hamiltonian then yields
\beq E= -1/2 (\partial_i u_0)^2, \label{negener} \eeq
which can be made arbitrarily negative by an appropriate choice of
$u_0$.

Moreover, as Clayton points out, the negative energies are not
restricted to this special case. In particular, allowing $c_2 \ne
0$ does not affect the $\overrightarrow{P} \cdot
\overrightarrow{\partial} u_0$ term in the Hamiltonian and even
produces additional questionable terms. The indefinite nature of
the decoupled Hamiltonian contrasts with with the wave energy
densities of Section \ref{Linear}, which clearly can be made
positive definite in the Maxwell-like case. The key point is that
the wave results are in the linearized theory and associated with
quadratic parts of the Hamiltonian, while the indefinite terms
appear at higher orders. For example, the linearized constraint
equation $v^0 = 0$ eliminates $\overrightarrow{P} \cdot
\overrightarrow{\partial} u_0$ from the Maxwell-like Hamiltonian
and forces the $u_0$ in (\ref{negener}) to be quadratic or higher
in the perturbation. Thus, the indefinite pieces begin to appear
at quartic order in the Hamiltonian. This indefiniteness at higher
orders implies that the decoupled, linearized results of Lim and
the ``coupled", linearized analysis of this paper generally do not
detect possible energies of arbitrary sign in the fully non-linear
decoupled Einstein-Aether theory.

\section{Discussion}

In this paper we have derived two energy-momentum pseudotensor
expressions for the Einstein-Aether theory and used them to
compute the energy densities of weak gravitational waves and the
total energy of an asymptotically flat solution. The constraints
of Section \ref{Linear} show that a sector of this LV model
satisfies the important theoretical condition of positive energy
in the linearized case. However, a remaining open question is
whether the energy remains positive when we consider the full
non-linear theory. We have argued that in the decoupled limit the
$c_1 \ne 0 $ non-linear sigma model is immune to the sickness of
energies of indefinite sign. However, other special cases of the
coupling constants yield negative energy solutions even when the
linearized theory has positive energy. A complete answer to the
question of positivity of energy in the non-linear theory is not
yet in hand.

The recipe for the Weinberg pseudotensor discussed in Section
\ref{Weinberg} and the Einstein superpotential and pseudotensor
derived in Section \ref{Einstein} also has applications in
studying the emission of gravitational-aether radiation from
astrophysical sources. In \cite{21} the analog of the quadrupole
formula (which also involves monopole and dipole moments), is
obtained using a pseudotensor expression derived from the related
Noether charge approach. This expression is then used to track
radiative energy losses and study constraints on the model.

\section*{Acknowledgements}

The author thanks Ted Jacobson for valuable discussions and
editing advice and Albert Roura for comments on an earlier draft.
This work was supported in part by the NSF under grant PHY-0300710
at the University of Maryland.

\section*{Appendix}\label{Appendix}
\renewcommand{\theequation}{A.\arabic{equation}}
\setcounter{equation}{0}

\subsection*{Background}
\label{background}

In this appendix we will derive the Einstein psuedotensor and
superpotential using the Noether current formalism of Julia and
Silva \cite{18} applied to Lagrangians that depend on the fields
and their first and second derivatives. We can write a variation
in the Lagrangian as
\beq \delta L = \frac{\partial L}{\partial \psi} \delta
\psi+\frac{\partial L}{\partial(\partial_\mu \psi)}
\delta(\partial_\mu \psi)+\frac{\partial L}{\partial(\partial_\mu
\partial_\nu \psi)} \delta(\partial_\mu
\partial_\nu \psi) \label{var} \eeq
and then integrate by parts to isolate the equations of motion $E$
and a symplectic current $\theta^\mu$,
\beq \delta L = E \delta \psi + \partial_\mu \theta^\mu,
 \eeq
where
\bea \partial_\mu \theta^\mu &=& \partial_\mu \left(\frac{\partial
L}{\partial(\partial_\mu \psi)}\delta
\psi-\partial_\nu(\frac{\partial L}{\partial(\partial_\mu
\partial_\nu \psi)})\delta \psi\right. \nonumber\\&& \left.+ \frac{\partial L}{\partial(\partial_\mu \partial_\nu \psi)}\partial_\nu(\delta
\psi)\right). \label{cons1}\eea
If the action associated with $L$ is invariant under a continuous
transformation of the fields, $\delta L = \partial_\mu S^\mu$.
Thus, we have the equation
\beq \partial_\m (S^\mu-\theta^\mu) = E \delta \psi.
\label{varL}\eeq
This identifies the on-shell ($E=0$) conserved Noether current,
\bea J^u &=& \theta^\mu-S^\mu = \frac{\partial
L}{\partial(\partial_\mu \psi)}\delta
\psi-\partial_\nu(\frac{\partial L}{\partial(\partial_\mu
\partial_\nu \psi)})\delta \psi \nonumber\\&&+ \frac{\partial L}
{\partial(\partial_\mu \partial_\nu \psi)}\partial_\nu(\delta
\psi)-S^\mu. \label{consvcurrent} \eea
We now want to consider a gauge transformation of the fields that
involves derivatives of the generator $\xi^A(x)$. Here we will
focus on the special case restricting attention to only the first
derivative. Following the analysis and notation of \cite{18} we
parameterize the gauge transformation as
\beq \delta \psi = \xi^A \Delta_A + (\partial_\nu \xi^A)
\Delta^\nu_{A} \label{deltapsi}\eeq
where $A$ is an internal or spacetime index and $\Delta$ is a
transformation matrix. The quantity $S^\mu$ can be expressed
similarly as
\beq S^\mu = \xi^A \Sigma^{\mu}_{A}+(\partial_\nu \xi^A)
\Sigma^{\mu \nu}_{A}+(\partial_\tau \partial_\nu \xi^A)
\Sigma^{\mu(\tau \nu)}_{A}.\label{surface}\eeq
Inserting these forms into (\ref{consvcurrent}) and combining
terms, we find that on-shell
\beq \partial_\mu\left(\xi^A J^\mu_{A}+\partial_\nu \xi^A U^{\mu
\nu}_{A}+\partial_\tau \partial_\nu \xi^A V^{\mu(\nu
\tau)}_{A}\right) = 0, \label{localcons} \eeq
where
\begin{eqnarray}
J^{\mu}_{A} &=&  \left(\frac{\partial L}{\partial(\partial_\mu
\psi)}-\partial_\nu(\frac{\partial L}{\partial(\partial_\mu
\partial_\nu \psi)})\right) \Delta_{A}
\nonumber\\&&+\left(\frac{\partial L}{\partial(\partial_\mu \partial_\nu \psi)}\right)\partial_\nu \Delta_{A}-\S^{\mu}_A \label{Jeqn}\\
U^{\mu\nu}_{A} &=&  \left(\frac{\partial L}{\partial(\partial_\mu
\psi)}-\partial_\tau(\frac{\partial L}{\partial(\partial_\mu
\partial_\tau \psi)})\right)
\Delta^{\nu}_{A}\nonumber\\
&&+\left(\frac{\partial L}{\partial(\partial_\mu
\partial_\tau \psi)}\right)\partial_\tau \Delta^\nu_{A}
+\frac{\partial L}{\partial(\partial_\mu \partial_\nu
\psi)}\Delta_A\nonumber\\&&
-\S^{\mu\nu}_A  \label{Ueqn}\\
V^{\mu(\tau\nu)}_{A} &=& \frac{\partial L}{\partial(\partial_\mu
\partial_\tau \psi)} \Delta^{\nu}_{A}-\Sigma^{\mu(\tau \nu)}_{A}.
\label{Veqn}
\end{eqnarray}
Since $\xi^A$ and its derivatives should be arbitrary and
independent, this single equation decomposes into 4 equations
\begin{eqnarray}
\partial_\mu J^\mu_{A} &\approx& 0 \label{cascade1}\\
J^\mu_{A}+\partial_\nu U^{\nu\mu}_{A} &\approx& 0 \label{cascade2}\\
U^{(\mu\nu)}_{A}+\partial_\tau V^{\tau(\mu \nu)}_{A} &=& 0 \label{cascade3}\\
V^{(\mu\nu\tau)}_{A} &=& 0\label{cascade4}.
\end{eqnarray}
The first two equations hold on-shell, while the last two are
identities (since there are no second or third derivatives of
$\xi^A$ on the right hand side of (\ref{cons1})).

The gauge symmetry implies that $J^\mu_{A}$ is conserved and equal
to the divergence of the superpotential $U^{\mu\nu}_{A}$. Since
$\xi^A = \xi^{A}(x)$, the Noether current $J^{\mu}$
(\ref{consvcurrent}) will now be parameter dependent. Let us
consider a one parameter subgroup of the local gauge or
diffeomorphism symmetry where $\xi^A$ has the decomposition,
\beq \xi^{A}(x) = \epsilon(x) \xi^{A}_0, \label{Jpsi} \eeq
and $\xi^{A}_0$ is fixed. Inserting this form into
(\ref{cascade2}) produces
\beq \xi^{A}_0 J^{\mu}_{A}+\partial_{\nu}(\xi^{A}_0
U^{\nu\mu}_{A}) = 0, \label{generalcons} \eeq
with $J^{\mu}_{\xi_0} = \xi^{A}_0 J^{\mu}_{A}$. The conserved
charge is
\beq Q = \int J^{0}_{\xi_0} d^3 x = \oint \xi^{A}_0 U^{\nu 0}_{A}
n_\nu d^2 x \eeq
$Q$ depends on the choice of $\xi^A_0$ and can be expressed in
terms of the gauge fields up using (\ref{Ueqn}). If $\xi^A_0$ is
an asymptotic translation in an asymptotically flat spacetime,
then the conserved charge will be a total energy or momentum.
Using the variational principle ($\delta S = 0 \Rightarrow$
equations of motion), we will show in the next section that the
choice of $\xi^A_0$ is subject to certain boundary conditions at
infinity. We now apply this type of Lagrangian analysis to the
Einstein-Aether theory.

\subsection*{Application}

Assume that the Lagrangian density ${\cal L}(\psi,\partial
\psi,\partial^{2}\psi)$ is invariant under diffeomorphisms and is
a combination of a scalar density $\widetilde{{\cal L}}$ and a
total divergence $\partial_\m W^{\m}$,
\beq {\cal L}(\psi,\partial \psi,\partial^{2} \psi) =
\widetilde{{\cal L}}(\psi,\partial \psi,\partial^{2}
\psi)+\partial_\m [W^{\mu}(\psi,\partial \psi)].
\label{genaction}\eeq
If $W^\m$ is a vector density then the total divergence is a
scalar density, but we allow for a non-covariant total divergence.
For a variation that is an infinitesimal diffeomorphism generated
by a vector field $\xi^\mu$, we have $\delta \widetilde{{\cal L}}
=
\partial_\mu(\xi^\mu \widetilde{{\cal L}})$ since $\widetilde{{\cal L}}$
is a scalar density and $\delta(\partial_\mu
W^\mu)=\partial_\mu(\delta W^\mu)$. Therefore the surface term
$S^\mu$ in \ref{varL} has the form
\beq S^\mu = \xi^\mu \widetilde{{\cal L}} + \delta W^\mu.
\label{Seqn}  \eeq
Now consider the Einstein-Hilbert action
\beq S_{\rm EH} = \int \sqrt{-g}~ R \eeq
of pure GR. The Ricci scalar $R$ has a dependence on second
derivatives of the metric. In light of this, Einstein exploited a
property of the Hilbert action that allows it to be separated into
a bulk and a surface term
\beq \int \sqrt{-g}~ R ~d^4x = \int \sqrt{-g}L_{\rm
bulk}+\partial_{\mu}V^{\mu} d^4x. \eeq
This decomposition of the Ricci scalar takes the following form
\begin{eqnarray}
L_{\rm bulk} &=&
g^{\alpha\beta}\{\Gamma^{\eta}_{\alpha\delta}\Gamma^{\delta}_{\eta\beta}
-\Gamma^{\eta}_{\eta\delta}\Gamma^{\delta}_{\alpha\beta}\} \\
V^{\mu} &=& \sqrt{-g}\{\Gamma^{\mu}_{\alpha\beta}g^{\alpha\beta}-
\Gamma^{\beta}_{\alpha\beta}g^{\mu\alpha}\}.
\end{eqnarray}
where $\Gamma$ is the Levi-Civita connection. One can eliminate
the total divergence by adding its negative to the
Einstein-Hilbert action
\beq \int \sqrt{-g}L_{\rm bulk} = \int \sqrt{-g}~ R ~d^4x -
\partial_{\mu}V^{\mu} d^4x. \eeq
The elimination does not affect the equations of motion and is
consistent with the general action (\ref{genaction}) with
$\widetilde{{\cal L}} = \sqrt{-g}~R$ and $W^\mu=-V^\mu$. The
result of this is a loss of diffeomorphism invariance since the
remaining $L_{\rm bulk}$ in the ``$\Gamma^2$" action is not a
scalar. We have allowed for this possibility with the
non-covariant $\partial_\mu W^\mu$ term in (\ref{genaction}).

With ${\cal L}(\psi,\partial \psi,\partial^{2}\psi)$ the bulk part
of the Einstein-Hilbert action plus the the aether terms in
(\ref{action}), we arrive at the Einstein-Aether form of
(\ref{varL}) on shell
\beq \partial_\mu \left(\frac{\partial
\cal{L}}{\partial(\partial_\mu g_{\alpha\beta})}\delta
g_{\alpha\beta}+\frac{\partial \cal{L}}{\partial(\partial_\mu
u^{\alpha})} \delta u^{\alpha} -S^\mu\right) = 0. \label{consg}
\eeq
The second derivative terms in (\ref{consvcurrent}) vanish in this
case. Under a diffeomorphism generated by a vector field $\xi^\nu$
the variation of the metric and the aether is simply the Lie
derivative, $\delta g_{\alpha\beta} = \xi^\nu\partial_\nu
g_{\alpha\beta}+
\partial_\alpha \xi^\nu g_{\nu\beta} +
\partial_\beta \xi^\nu g_{\alpha\nu}$ and $\delta u^{\alpha} =
\xi^{\gamma}\partial_{\gamma} u^{\alpha} -
u^{\gamma}\partial_{\gamma}\xi^{\alpha}$. It follows from
(\ref{Seqn}) that $S^\mu = \xi^\mu
L_{bulk}-\sqrt{-g}(\partial_\delta
\partial_\nu \xi^\mu g^{\delta\nu}-\partial_\delta \partial_\nu \xi^\nu
g^{\mu\delta})$. Inserting these forms into (\ref{consg}) produces
\begin{eqnarray}
t_{\nu}{}^\mu &=& \sqrt{-g}\left(\frac{\partial
L}{\partial(\partial_\mu g_{\alpha\beta})} \partial_\nu
g_{\alpha\beta} + \frac{\partial L}{\partial(\partial_\mu
u^{\alpha})} \partial_{\nu} u^{\alpha}\right. \nonumber\\&& \left.
- \delta^\mu_\nu
L\right)\\
U_{\nu}{}^{\mu\gamma} &=& \sqrt{-g}\left(\frac{\partial
L}{\partial(\partial_\mu
g_{\a\b})}(\delta^{\gamma}_{\alpha}g_{\nu\beta}+\delta^{\gamma}_{\beta}g_{\nu\alpha})\right.
\nonumber\\&& \left.-\frac{\partial L}{\partial(\partial_\mu
u^{\alpha})} \delta^{\alpha}_{\nu} u^{\gamma}\right)
\\
V_{\nu}{}^{\mu(\gamma\lambda)} &=& \sqrt{-g}\left(
\delta^{\mu}_{\nu}g^{\lambda\gamma}-\frac{1}{2}\delta^{\gamma}_{\nu}g^{\lambda\mu}-\frac{1}{2}\delta^{\lambda}_{\nu}g^{\gamma\mu}
\right)
\end{eqnarray}
as coefficients of $\xi^{\mu}$, $\partial \xi^{\mu}$ and
$\partial^2 \xi^{\mu}$ respectively. $t_{\nu}{}^{\mu}$ and
$U_{\nu}{}^{\mu\gamma}$, and $V_{\nu}{}^{\mu(\gamma\lambda)}$ are
the analogs of $J^{\mu}_{A}$, $U^{\mu\nu}_{A}$,
$V^{\mu(\gamma\lambda)}_{A}$ in (\ref{Jeqn})-(\ref{Veqn}). The
resulting equations due to the arbitrariness and independence of
the derivatives of $\xi^{\mu}$ are
\begin{eqnarray}
\partial_\mu t_\nu{}^\mu &\approx& 0\\
t_\nu{}^\mu &\approx& -\partial_\gamma U_\nu{}^{\gamma\mu}\\
U_{\nu}{}^{(\gamma\mu)} + \partial_\lambda
V_{\nu}{}^{\lambda(\mu\gamma)} &=& 0\\
V_{\nu}{}^{(\lambda\nu\gamma)} &=& 0.
\end{eqnarray}
Following (\ref{Jpsi}), we can keep the $\xi^\nu$ vector fixed
(and determine it later for each conserved charge) by choosing
$\xi^\nu = \e(x) \xi^\nu_0$. The main result, as before, is
\beq \xi^\nu_{0} t_{\nu}{}^{\mu} = -\partial_\gamma(\xi^\nu_{0}
U_{\nu}{}^{\gamma\mu}) \label{Eeqn}\eeq
showing that a Noether charge is again obtained as a surface term.
Einstein effectively chose the $\xi^\nu_{0}$ vector to be a
constant in (\ref{Eeqn}), reducing the pseudotensor to a form
consistent with the flat spacetime canonical stress tensor
(\ref{flatcan}). However, this choice is not inconsequential. The
variational principle for the $\Gamma^2$ action requires the
vanishing of the surface term in the asymptotic region,
\beq \int_{S_{\infty}} \frac{{\partial \cal
L}}{\partial(\partial_\mu g_{\a \b})} \delta g_{\a \b} +
\frac{{\partial \cal L}}{\partial(\partial_\mu u^{\a})} \delta
u^\a . \eeq
Therefore we have Dirichlet boundary conditions $\delta
g_{\a\b}=0$ and $\delta u^\a = 0$ on the metric and the aether at
infinity. Inserting the Lie derivatives for the variations above,
we see that as $r\rightarrow \infty$
\begin{eqnarray}
\nabla_{(\a} \xi_{\b)} \rightarrow 0\\
{\cal L}_{_\xi} u^{\a} \rightarrow 0.
\end{eqnarray}
Since $\xi^\nu$ has been chosen to be constant everywhere and
$u^\a$ is asymptotically constant, the connection coefficients
must vanish as one approaches spatial infinity. Thus, one must
compute the pseudotensor and superpotential in a coordinate system
where the connection vanishes asymptotically as $O(1/r)$ or
faster.

\end{document}